\RequirePackage[hyphens]{url}
\documentclass[conference]{IEEEtran}
\usepackage{fancyhdr}
\IEEEoverridecommandlockouts
\usepackage{cite}

\usepackage[subtle,tracking=normal]{savetrees}

\usepackage{multicol}
\usepackage{lipsum}

\usepackage{hyperref}

\usepackage{soul}
\usepackage{color}

\usepackage{amsmath}
\usepackage{graphicx}

\ifCLASSINFOpdf
\else
\fi
\usepackage{array}
\usepackage{booktabs}

\begin{document}
%
\title{CATCH: a Cost Analysis Tool for Co-optimization of chiplet-based Heterogeneous systems}



\author{
\IEEEauthorblockN{Alexander Graening\textsuperscript{*}, Jonti Talukdar\textsuperscript{**}, Saptadeep Pal\textsuperscript{\textdagger}, Krishnendu Chakrabarty\textsuperscript{\textdagger\textdagger}, Puneet Gupta\textsuperscript{*}}
\thanks{This work was supported in part by CHIMES, one of the seven centers in JUMP 2.0, a Semiconductor Research Corporation (SRC) program sponsored by DARPA, ASML and the CDEN center.}
\\
\begin{minipage}[t]{0.5\textwidth}
\centering
\textsuperscript{*}Department of Electrical and Computer Engineering\\
University of California, Los Angeles\\
agraening@ucla.edu, puneet@ee.ucla.edu
\end{minipage}%
\begin{minipage}[t]{0.5\textwidth}
\centering
\textsuperscript{**}Department of Electrical and Computer Engineering\\
Duke University\\
jonti.talukdar@duke.edu
\end{minipage}%
\\
\\
\begin{minipage}[t]{0.5\textwidth}
\centering
\textsuperscript{\textdagger}Etched\\
saptadeep@etched.com
\end{minipage}
\begin{minipage}[t]{0.5\textwidth}
\centering
\textsuperscript{\textdagger\textdagger}School of Electrical, Computer and Energy Engineering\\
Arizona State University\\
krishnendu.chakrabarty@asu.edu
\end{minipage}
}



\maketitle

\begin{abstract}

With the increasing prevalence of chiplet systems in high-performance computing applications, the number of design options has increased dramatically. Instead of chips defaulting to a single die design, now there are options for 2.5D and 3D stacking along with a plethora of choices regarding configurations and processes. For chiplet-based designs, high-impact decisions such as those regarding the number of chiplets, the design partitions, the interconnect types, and other factors must be made early in the development process. In this work, we describe an open-source tool, CATCH, that can be used to guide these early design choices. We also present case studies showing some of the insights we can draw by using this tool. We look at case studies on optimal chip size, defect density, test cost, IO types, assembly processes, and substrates.


\end{abstract}


%


\section{Introduction}

    Chiplets are becoming increasingly important in modern chip design. This is partly due to the slowing of Moore's Law scaling. In order to continue scaling the number of transistors in a package, it is necessary to increase the size of the package. As simply making an arbitrarily large die runs into reticle stitching complications and low die yield, chiplet systems are often used to integrate multiple dies in a package. This can include both 2.5D and 3D stacking. NVIDIA's advanced GPU designs such as the A100 \cite{nvidiaA100} are good examples of this strategy. The NVIDIA A100 integrates high bandwidth memory (HBM), which is a form of 3D stacked memory alongside a large GPU die. The A100 contains a single GPU die along with 6 HBM stacks. There are other examples of chiplets in industry from both Intel \cite{ponteVecchio} and AMD \cite{amdInstinct}.\par

    \subsection{Chipletization Size Regimes}

        Each of the previous examples of chiplet systems are larger than standard reticle size. These systems would be impossible or impractical to implement as a single die. Although this is the dominant current application for chiplets, there are many reasons to use chiplets for smaller systems.\par
    
        Chiplet systems enable integrating multiple process technologies in a package. This can allow for implementing different parts of the design in more favorable technologies. For example, some analog designs have better properties in less-advanced technology nodes. Also, GaN is better than silicon for power circuits, but not as good as silicon for logic \cite{Jones2016}. Chiplet systems can mix these technologies in a package.\par
    
        Another benefit of chiplets is the possibility of reducing cost by moving from a monolithic to a chiplet system. For large systems, splitting a large die into smaller dies can increase yield and decrease the corresponding cost. For smaller systems, the yield benefit is less significant, but reuse of chiplets can amortize non-recurring engineering (NRE) costs over larger effective volumes. These cost benefits are balanced against the added expenses of advanced packaging. To make an educated decision regarding a design's chiplet configuration, it is important to conduct detailed analysis of the costs associated with chipletization.\par

    \subsection{Variety of Configuration Options for Chiplet Systems}

        Chiplets provide a wide range of configuration options. These include the number of chiplets and the stacking configuration. Should a design be a single die? Should it be split and integrated on an interposer? Should it be 3D stacked? Each option has its own cost and yield implications. Additionally, integration substrates, bonding pitch, IO cell design, testing process, and many other factors all influence the cost.\par

    \subsection{Need for a Chiplet Cost Modeling Tool}

        The design choices for chiplet configuration options need to be made early in the design process. For example, splitting a monolithic design into two 2.5D integrated chiplets will require adding IO cells and reorganizing the design blocks for each of the component chiplets. It will require designing an integration substrate and ensuring that signals crossing the chiplet boundary can handle any latency added for the inter-chiplet communication. Switching from a monolithic to a chiplet-based design late in the design process can result in delays and increased design costs.\par

        This paper describes CATCH, a tool to evaluate the cost impacts of a wide range of chiplet-specific design choices. While this work specifically addresses cost modeling, it is also important to evaluate factors such as performance, thermal, and power. Therefore, our cost-modeling tool is designed to streamline integration with other tools for co-optimization in future work.\par


    \subsection{Prior Work}

        A number of previous works have addressed parts of this problem. Our modeling tool is an extended version of the model originally introduced in \cite{Graening2023}. This paper used an early version of our model that lacked support for NRE cost, test cost, wafer utilization, IO reach, and multiple substrate types. We greatly expand on this model in the current work.\par

        Chiplet Actuary \cite{Feng2022} is an open-source cost modeling tool that addresses many of the same problems we are trying to address. This tool thoroughly addresses NRE costs, but lacks the flexibility of our tool to model arbitrary stacking configurations of dies and has less detailed modeling of recurring costs.\par

        A much more complete cost model was described in \cite{Ahmad2022}. It describes a detailed spreadsheet-based cost model that covers die cost, die yield, assembly yield, testing cost, yield/cost propagation during assembly, IO cost, and NRE costs. In addition to covering these costs, the authors address some factors that we do not address in our work. These include the cost impacts of adding redundancy and the impacts of changing cost and yield over time as well as the impact of uncertainty on the assumed parameters. However, \cite{Ahmad2022} does not include assembly cost as an additional cost due to time spent on a bonding machine for a complicated chiplet system. Additionally, there is no general IO model based on a netlist and no sense of how reach or number of IOs will impact the size of a die with a certain bonding pitch. The system modeling is also more limited for co-optimization studies as the focus is on cost.\par


        Several works that do not focus on cost modeling use simplified equations to calculate the cost of partitioning a design into various chiplet configurations as in \cite{Ehrett2021, Li2024, Chen2024}. Note that \cite{Ehrett2021} depends on reuse, which is calculated from an equation that incorrectly models this as a direct function of area where smaller chiplets are more reusable. In reality, larger chiplets with more features can be more reusable than smaller chiplets as they can be used for a wider range of designs if you turn off the unnecessary blocks. As such, we avoid including a simple reuse model as it is important to know a full design portfolio to assess the impacts of reuse. If reuse is known, this can be accounted for in our model by adjusting die quantities for a chiplet assembly.\par

        A number of other works also use a simplified cost model as part of a larger design space exploration either for chiplets or a specific aspect of chiplet design. \cite{Pal2020, Stow2017, Kim2020, Stow2019, Hao2023}\par


    \subsection{Overview}

        Our model can be broken conceptually into two parts. The first is the generic descriptive system model and the second is the set of cost modeling functions that are used to generate the system cost. For the generic system model, we ensure assumptions about IO placement, die sizes, power, and power pins are met without requiring explicit user attention. The cost functions then operate on the computed parameters of the system described in the system definition.\par
        
        In addition to providing a tool, we present interesting studies for the following design factors:\par

        \begin{itemize}
            \item Cost Optimal Chiplet Size
            \item Testing Cost
            \item IO Reach
            \item Assembly Costs
            \item Interposer Type
            \item NRE cost
        \end{itemize}

        This paper starts with a description of the system model and is followed by a detailed description of the cost model functions. After this, we present a description of the input formats along with comments on the sources of the numbers used for our studies. This is followed by the case studies section with associated discussions to show how the tool can be used for design space exploration. Our cost modeling tool is open source at \href{https://github.com/nanocad-lab/cost_model_chiplets}{https://github.com/nanocad-lab/cost\_model\_chiplets} along with a script to allow for easy reproduction of all the plots in this paper.\par



\section{Modeling Tool Overview}

One advantage of our model compared to other cost models is that we attempt to capture the features of an arbitrary design that ensures consistency between factors such as area, power, and IOs. This is useful for modeling the cost of a specific system, running a design space exploration where you want all designs to remain feasible while you sweep a parameter, and for integrations with other tools that require the calculated non-cost parameters generated in the model.\par

    \subsection{Supported Physical Architectures}

\begin{figure}[t]
            \centering
            \includegraphics[width=\columnwidth]{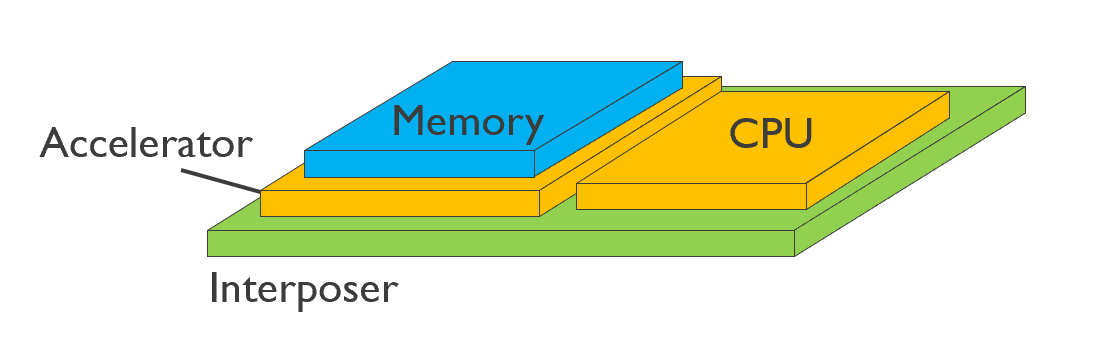}
            \caption{Supported Structures. The tool can model any arbitrary 2.5D or 3D stack of \textit{Chip} objects provided that a larger die is not placed on top of multiple smaller dies.}
            \label{fig:supported_structures}
        \end{figure}

        The modeling tool is centered around the definition of \textit{Chip} objects. These \textit{Chip} objects have a variety of parameters some of which are directly user-defined and others that are calculated from the provided parameters. Architectures are modeled as stacks of \textit{Chips} with different parameters. This means that a PCB with some chips on top could be modeled just as well as a 2.5D integrated system with multiple chiplets in a single package on an interposer or a 3D system where we have multiple dies stacked on top of each other vertically. As shown in Figure \ref{fig:supported_structures}, combinations of stacked \textit{Chips} are supported so long as top \textit{Chips} are smaller than lower \textit{chips}. By changing parameters, it is possible to define a \textit{Chip} object with low cost per unit area to represent a PCB. With a different set of parameters, a \textit{Chip} could be a passive interposer, PCB, logic, memory, or analog in any technology node.\par

    \subsection{Recursive Structure}

        CATCH is designed to facilitate reuse of functions and support arbitrary stacking. Due to this, the structure of the system is defined in the \textit{Chip} definitions and the same functions are run on each \textit{Chip} object recursively.\par

        Each \textit{Chip} object contains the information specific to the die, substrate, or PCB represented along with a list of all \textit{Chip} objects that are stacked on top of it. There are no arbitrary limits to this structure, so a \textit{Chip} may contain a list of as many stacked \textit{Chip} objects as desired and each of the stacked chips can contain its own additional stacked \textit{Chip} objects. In the example in Figure \ref{fig:supported_structures}, the definition of the Interposer \textit{Chip} object contains the definitions of both the CPU and the Accelerator \textit{Chip} objects. The Memory \textit{Chip} object is then nested in the definition for the Accelerator \textit{Chip}. \par

    \subsection{Area Propagation}

        Area is computed as the largest of 3 values: area of the core plus IO cells, area of the \textit{Chips} stacked on top of the \textit{Chip}, or area required by the pads. Note that we also include a ``black box area'' parameter that allows overriding this computation and setting an exact area for each \textit{Chip}.\par

        \begin{equation}
            A_{chip}=\max{\{A_{core}+A_{IO},A_{stack},A_{pads}\}}
        \end{equation}

        \subsubsection{Core Area}

            Core area ($A_{core}$) is simply the core area defined in the system definition. Note that a \textit{Chip} such as an interposer or PCB will not have a core area since those areas are determined by the areas of the bonding pads or stacked dies. \par

        \subsubsection{IO Area} \label{sec:IO_area}

            IO area ($A_{IO}$) is the area of the IO cells required according to the bandwidth requirements and IO type in the netlist. The IO cell area is only added to the terminal \textit{Chip} objects on a net. If the interconnect passes through another die or interposer, this will add to the number of bumps but not to the IO cell area for that \textit{Chip}. \par

            IO area is calculated by adding up the number of nets that cross the chip boundary for each chip object. Since the design netlist assigns an IO type and bandwidth to each connection, the model computes the number of IO cells required to satisfy the bandwidth requirement for the connection and sums up the total area of the IO cells in the chip. It is possible for the transmit IO cell to have a different size than the receive IO cell for unidirectional IO types. To capture this, the IO cells define both a transmit and a receive area. The appropriate area is added for each connection based on the netlist direction.\par

            The netlist is stored in a series of adjacency matrices. Each IO type in the design has its own adjacency matrix where the row indicates the TX chip and the column indicates the RX chip. To calculate the area required by the IO cells, the model iterates through each IO type that has an adjacency matrix in the design and takes the sum of the entries in the row and column corresponding to the chip:\par

            \begin{equation}
                A_{IO} = \sum_{i=1}^n A_{IO,i}
            \end{equation}

            \noindent where $n$ is the number of IO cells in the design.

            \begin{equation}
                A_{IO,i} = a_{i,TX}\sum M_i[b,:] + a_{i,RX}\sum M_i[:,b]
            \end{equation}

            Note that $M_i$ is the adjacency matrix for IO cell $i$. $a_{i,TX}$ and $a_{i,RX}$ are the transmit and receiving area of the IO cell respectively. Note that M is the adjacency matrix calculated such that the numbers indicate the number of instances of the IO cell required to satisfy the required bandwidth. Diagonal elements of the adjacency matrix are set to 0 to avoid double counting. The number for each of the other entries is calculated as below:\par

            \begin{equation}
                N_{IO\_instances} = \left\lceil\frac{B_{req}}{B_{IO}}\right\rceil
            \end{equation}

            \noindent where $B_{req}$ is the required bandwidth and $B_{IO}$ is the bandwidth for the specific IO.\par

        \subsubsection{Stack Area}

            The stacked \textit{Chip} area is simply a sum of the areas of each \textit{Chip} stacked on top of the \textit{Chip} in question:\par

            \begin{equation}
                A_{stack} = \sum_{i=0}^nA_{subchip,i}
            \end{equation}

            \noindent where $A_{subchip}$ is the area of the subchip expanded by half of the die separation distance on each edge along with an edge exclusion zone the defines the distance a die may be placed from the edge of the lower chip when stacking. This is due to the minimum die separation for bonding. We make the simplifying assumption that the dies pack together well and that the total area required for bonding is the sum of the subchip areas.\par


        \subsubsection{IO Pad Area} \label{sec:io_pad_area}

\begin{figure}[t]
                \centering
                \includegraphics[width=0.8\columnwidth]{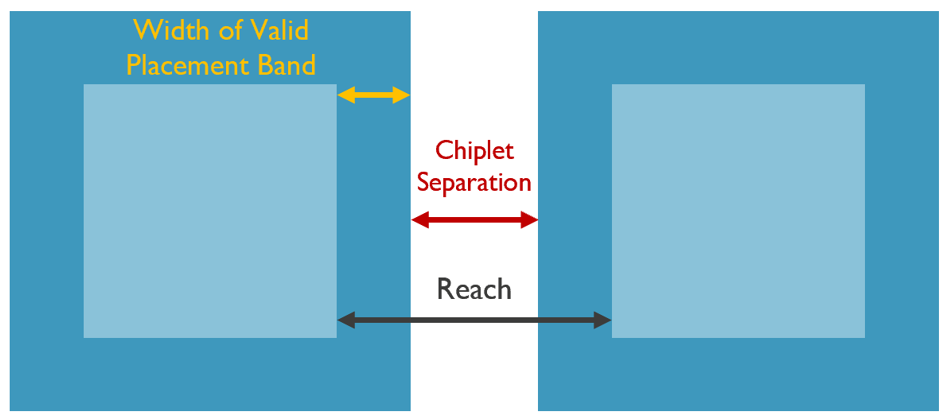}
                \caption{IO Pad Placement}
                \label{fig:io_pad_placement}
            \end{figure}

            Pad area is calculated based on the area required for the power and signal pads based on both the bump pitch and the reach of the IO cells for signal pads. Since each IO cell has a reach, the pads must be placed within a region at the edge of the die. Assuming nearest neighbor communication, this means that signal pads will have a valid placement region within a band with a thickness given by the following equation and shown in Figure \ref{fig:io_pad_placement}.\par

            \begin{equation}
                W_{placement\_band}=\frac{R-D_{chip\_separation}}{2}
            \end{equation}

\begin{figure}[t]
                \centering
                \includegraphics[width=0.5\columnwidth]{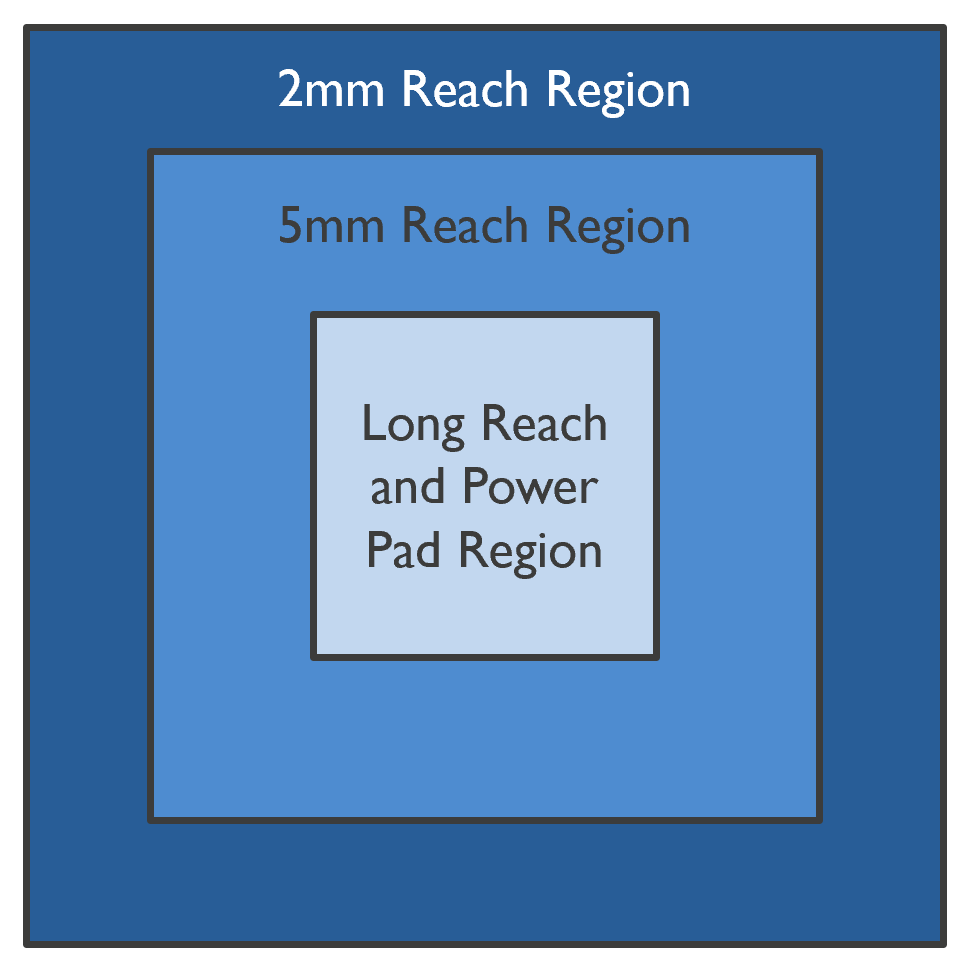}
                \caption{Multiple IO Type Pad Placement}
                \label{fig:multi_IO_pad_placement}
            \end{figure}

            To check this for multiple IO types, the IO cells are sorted according to reach and placed from the shortest reach to the longest reach starting from the outside of the die and moving inward. After all signal pads are placed, the power pads are placed with no reach restriction. If for any IO pad, there is not enough area that satisfies the reach constraint, the size of the chip is increased to increase the perimeter and thus the valid placement area before moving to the next IO type. Since the power pads do not have a reach constraint, they will only increase the chip area if there is not enough area at the center of the chip to place them. See Figure \ref{fig:multi_IO_pad_placement} for a general map of how these IOs are placed.\par

            There are several simplifying assumptions for the IO pad placement that impact the accuracy of the approximation for different situations. CATCH assumes neighbor to neighbor connections where each IO type has equal bandwidth connecting to chips in each direction from the chip. If a signal is expected to connect over a chip more than the minimum separation distance away, the valid placement region will be incorrect. If the chip primarily connects to a chip on one side, it will also be incorrect since the IOs are placed all around the chip perimeter.\par

            In addition to IO pad area, power pads and test pads must be placed. The power pad computation is described in section \ref{sec:power_pads} and the test pad computation is described below. Neither of these pad types have a reach constraint. \par



        \subsubsection{Test Pad Area} \label{sec:test_pad}

            We assume a scan chain test. $N_{sc}$ is the number of scan chains, $N_{IOs\_per\_sc}$ is the number of IOs per scan chain, and $N_{test\_IO\_offset}$ is an offset for the number of IOs to account for signals such as a test enable signal. The computation for the number of test IOs ($N_{test\_IOs}$) is shown in equation \ref{eq:test_ios}.\par

            \begin{equation} \label{eq:test_ios}
                N_{test\_IOs}=N_{sc}\times N_{IOs\_per\_sc} + N_{test\_IO\_offset}
            \end{equation}

    \subsection{Power Propagation} \label{sec:power}

        Power is computed as the sum of the core power, the power of stacked chips, and the additional power required by the IOs. Note that we also offer a ``black box power'' parameter that overrides this computation to allow setting the exact power for a specific \textit{Chip}.\par

        \begin{equation}
            \label{eq:total_power}
            P=P_{core}+P_{stack}+P_{IO}
        \end{equation}

        \subsubsection{Core Power}

            Core power is defined for each \textit{Chip}. This is one number and as it is used to compute the number of required power and ground connections, this is the maximum power required by the die. \par

        \subsubsection{Stack Power}

            Due to the recursive structure, the total power for the chip is the sum of the self power and the stack power which is the sum of the total powers of each chip on top of the chip in question.\par

            \begin{equation}
                P_{stack} = \sum_{i=1}^n P_i
            \end{equation}

        \noindent where $P_{stack}$ is the stack power and $P_i$ is the power of the i-th chip stacked on the current chip. $n$ is the number of stacked chips.\par

        \subsubsection{IO Power}

            Similar to the computation of the IO requirements described in section \ref{sec:IO_area}, each IO type has a defined bandwidth dependent power and  each connection has an average bandwidth utilization. This is added to the total power computation used for power pads.\par

        \subsubsection{Power Pads} \label{sec:power_pads}

            This impacts the pad area in addition to the signal IOs and test pads. The power that can be supported per pad is computed by finding the maximum pad current according to the operating voltage and the maximum current density times the area of the pad assuming a circular pad with a diameter half of the bonding pitch.\par

            \begin{equation}
                P_{pad} = V_{core}J_{max}\pi (d_{bonding}/4)^2
            \end{equation}

            \noindent where $P_{pad}$ is the power per pad, $V_{core}$ is the core voltage, $J_{max}$ is the maximum current density for the type of bond, and $d_{bonding}$ is the bonding pitch.

            The number of power pads is calculated according to the maximum power per bump and the total power $P$ of the chip. Note that this is multiplied by 2 since both power and ground pins are required.\par

            \begin{equation}
                N_{ppads}=2 \left \lceil \frac{P}{P_{pad}} \right \rceil
            \end{equation}

            $N_{ppads}$ is the number of power pads, $P$ is the total power of the chip and stack, and $P_{pad}$ is the maximum power per pad.\par

    \subsection{Connectivity Structure}

        The system definition includes a netlist which describes the connections between \textit{Chip} objects. Each defined connection has a \textit{from} and a \textit{to} \textit{Chip} and provides an IO type. The connection also includes either a required bandwidth or a ``count'' of the required number of instances of the IO type.\par

        Since IO types may be defined as either bidirectional or unidirectional, the \textit{from} and \textit{to} blocks are interchangeable in the case of a connection defined with a bidirectional IO type. In the case where an IO type is unidirectional, connections must be defined in each direction separately.\par

        We consider a source-sink model in this definition and do not directly support an IO with a single source and multiple sinks or vice versa. Defining an IO of one of these configurations within the current framework requires defining a dummy, 0-area \textit{Chip} object and custom IO types with different \textit{to} and \textit{from} areas.\par

\section{Cost Model} \label{sec:cost_model}

    \subsection{Manufacturing/assembly}
    
        \subsubsection{Die Cost}

            The cost of the individual die is given below as $C_{self}$.
            
            \begin{equation}
                C_{self}=C_{RE,self}+C_{NRE,self}
            \end{equation}

            The cost is split into recurring cost ($C_{RE,self}$) and non-recurring cost ($C_{NRE,self}$). $C_{RE,self}$ is defined below in Equation \ref{eq:re_cost}. 

            \begin{equation}
                \label{eq:re_cost}
                C_{RE,self} = \frac{C_{die}+C_{test}}{Y_{tested}}
            \end{equation}

            $Y_{tested}$ is the percentage of dies that pass test. This is typically higher than $Y_{true}$, which is the number of dies that are good (independent of test quality) according to our yield model. $C_{NRE,self}$ is defined according to Equation \ref{eq:nre_cost}.

            The assembly cost and yield are added as shown in equation \ref{eq:final_cost} along with the sum of stacked dies. \par

            \begin{equation}
                \label{eq:final_cost}
                 C = C_{RE} + C_{NRE}
            \end{equation}

            \begin{equation}
                \label{eq:total_re_cost}
                C_{RE}=\frac{C_{asmb} + C_{test,asmb} + C_{self,RE} + \sum_{i=1}^n C_{RE,i}}{Y_{tested,asmb}}
            \end{equation}

            \begin{equation}
                \label{eq:total_nre_cost}
                C_{NRE} = C_{NRE,self} + \sum_{i=1}^n C_{NRE,i}
            \end{equation}

            Each $C_{RE,i}$ and $C_{NRE,i}$ are computed recursively the same way as $C_{RE}$ and $C_{NRE}$. $C_{NRE}$ is kept separate from $C_{RE}$ as the recurring costs scale with yield while the non-recurring costs do not.

\begin{figure}[t]
                \centering
                \includegraphics[width=0.6\columnwidth]{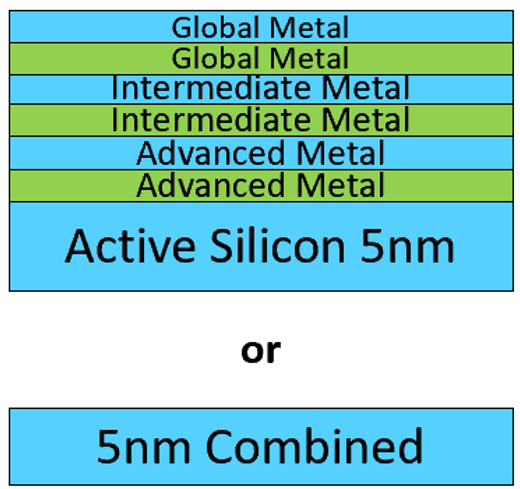}
                \caption{Stackup Options. It is possible to define the characteristics of a die either as a single \textit{Layer} or as a stack of \textit{Layers}.}
                \label{fig:stackup_options}
            \end{figure}

            Raw (untested) die cost is calculated according to the area of each die. The dies can be defined as a single ``layer'' or multiple layers which can have their own associated costs as shown in Figure \ref{fig:stackup_options}. In the case where there are multiple layers, the cost of the die is the sum of the costs of each layer.\par

            \begin{equation}
                \label{eq:multi_layer_cost_untested}
                 C_{die}=\sum_{i=1}^n C_{layer,i}
            \end{equation}
         
            The layer cost is calculated from the cost per $mm^2$. Since wafer costs for different technology nodes are publicly available, we compute the cost per $mm^2$ from the wafer cost. The cost also depends on the number of dies that fit on the wafer, so we estimate the number of dies that fit on the wafer to increase the cost per $mm^2$ according to the wasted silicon. We use the following method to determine how many dies will fit.\par

            Typical wafer blade dicing uses cuts that cross the full wafer. This means dies must be placed in a grid pattern. For a more flexible dicing method such as plasma dicing, it is possible to place the dies on the wafer more flexibly. This second case is more applicable if the die size is close to the reticle size since the reticle boundary can also limit flexibility of die placement. We support both options and recommend the grid option as default.\par
            
\begin{figure}[t]
                \centering
                \includegraphics[width=\columnwidth]{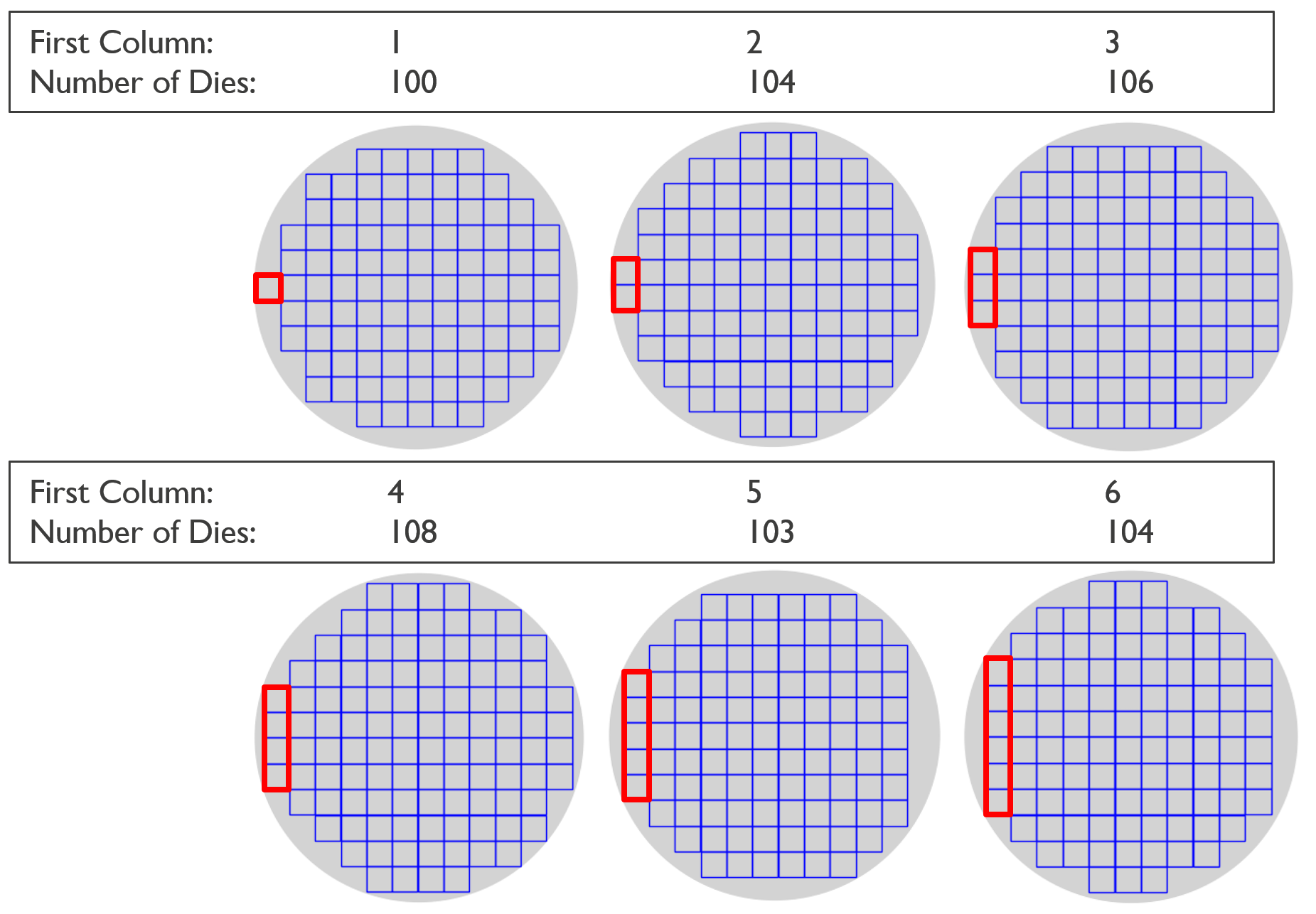}
                \caption{The die per wafer calculation will increase the number of dies in the first column iteratively and fit around that. The best of these will be selected as the solution. In this example, the best case was 4 dies in the first column with a total of 108 dies per wafer. This did not run past 6 dies in the first column as 7 would allow enough room for another column to the left and would look like one of the previous cases.}
                \label{fig:die_per_wafer}
            \end{figure}

            For the grid die placement, we use an iterative method to estimate the maximum number of dies that fit. We do this by increasing the number of dies in the first column of dies starting on the left starting with 1 until there is room for more dies to fit to the left of the column. This is shown in Figure \ref{fig:die_per_wafer}. The die calculation is similar to the no-grid option described below, but with the added constraint that all the dies must fall on the grid and a block of rows in accordance with the size of the first column are placed initially instead of a single row.\par

\begin{figure}[t]
                \centering
                \includegraphics[width=\columnwidth]{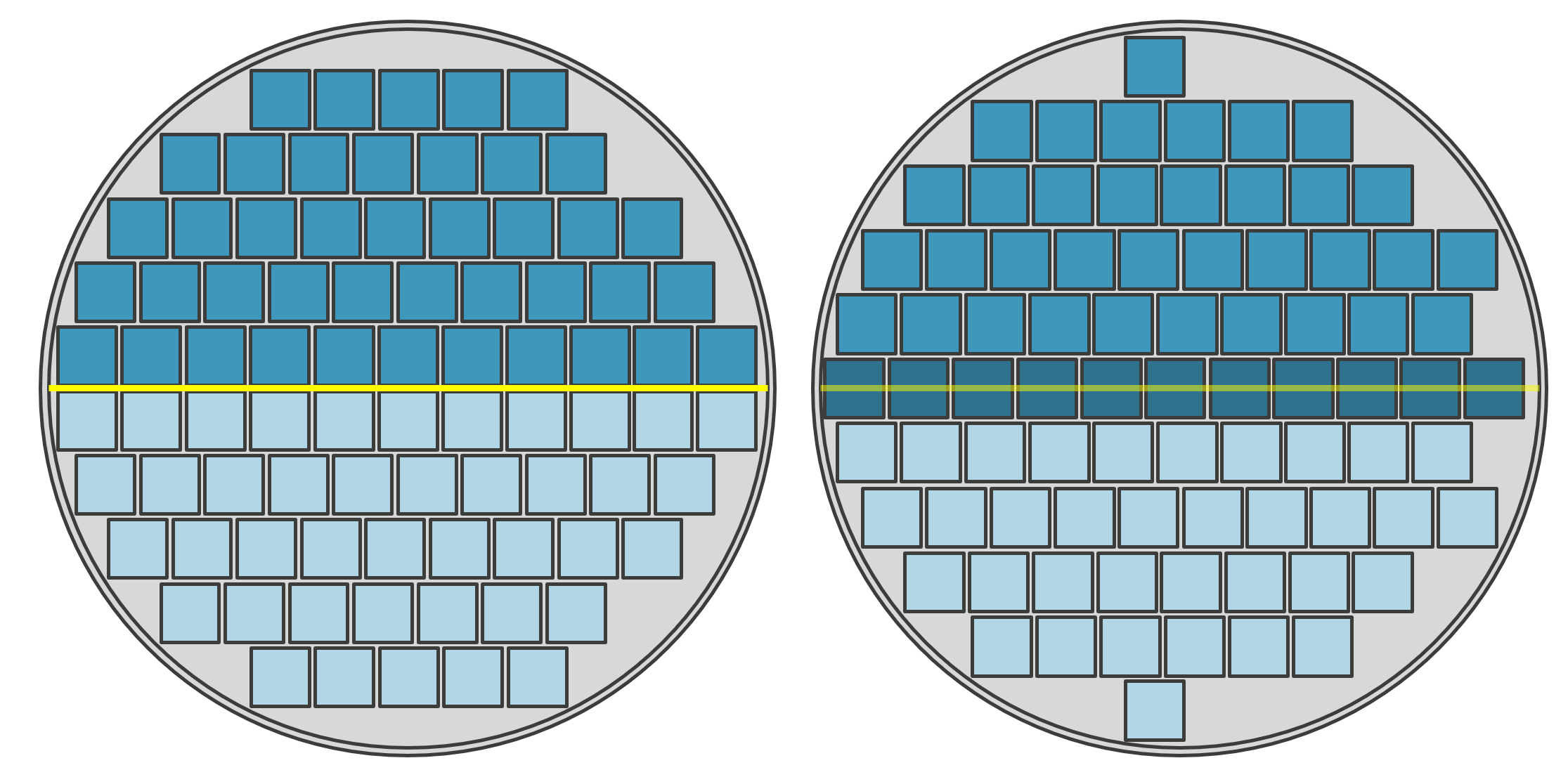}
                \caption{Die Per Wafer Calculation Without Grid. If using a dicing technique such as plasma dicing, there is less constraint that the dies are placed on a grid. We provide a flag to allow selecting this method.}
                \label{fig:die_per_wafer_nogrid}
            \end{figure}

            For the non-grid placement option, we only look at two cases. The first option places a row of chips above the diameter of the wafer as shown in Figure \ref{fig:die_per_wafer_nogrid} on the left. The number of chips ($N_0$) that fit here is shown in equation \ref{eq:count_above_diameter}.\par

            \begin{equation}
                \label{eq:count_above_diameter}
                N_0 = \left\lfloor\frac{\sqrt{r^2 - y_{chip}^2}}{x_{chip}}\right\rfloor
            \end{equation}

            \noindent where $r$ is the usable radius of the wafer (assuming some edge exclusion) and $x_{chip}$ and $y_{chip}$ are the are the x and y dimensions of the die.\par

            The second option places chips directly on the diameter of the wafer for the first row as shown in Figure \ref{fig:die_per_wafer_nogrid} on the right. The number of chips in the first placed row is given in equation \ref{eq:count_on_diameter}. The only difference from equation \ref{eq:count_above_diameter} is the height at which we calculate the length of the chord of the wafer.\par

            \begin{equation}
                \label{eq:count_on_diameter}
                N_0 = \left\lfloor\frac{\sqrt{r^2 - (y_{chip}/2)^2}}{x_{chip}}\right\rfloor
            \end{equation}

            In the first case, adding the first row above the diameter line of the wafer, additional rows of dies are added each above the previous until there is no room for an additional row and the result is doubled to account for the lower half of the wafer. In the second case the same is done except the first row is unique and is not counted twice along with the other rows.\par

            We include both scribe line and edge exclusion in this computation by simply reducing the size of the circle and increasing the size of the dies that are placed. Also, note that while the example figures show square dies, CATCH supports different aspect ratios.\par


            The cost per $mm^2$ is used to compute a cost per layer shown in equation \ref{eq:layer_cost_baseline}.\par

            \begin{equation}
                \label{eq:layer_cost_baseline}
                C_{layer\_baseline} = A_{chip}C_{per\_mm^2}
            \end{equation}
            
            A substantial percentage of costs come from time spent on the lithography machine \cite{DieSizeReticleLitho}. Since foundries typically place requirements on die size to ensure good reticle fit, we add an adjustment for the additional exposures required to fill a wafer with a poor reticle fit. We assume that a percentage of the wafer cost is due to time on the lithography machine and scale that cost according to the extra time spent on the machine when the chip is a poor fit for the reticle.\par

            \begin{equation}
                C_{layer} = C_{layer\_baseline}(1-P_{litho}+\frac{P_{litho}}{U_{reticle}})
            \end{equation}

\begin{figure}[t]
                \centering
                \includegraphics[width=\columnwidth]{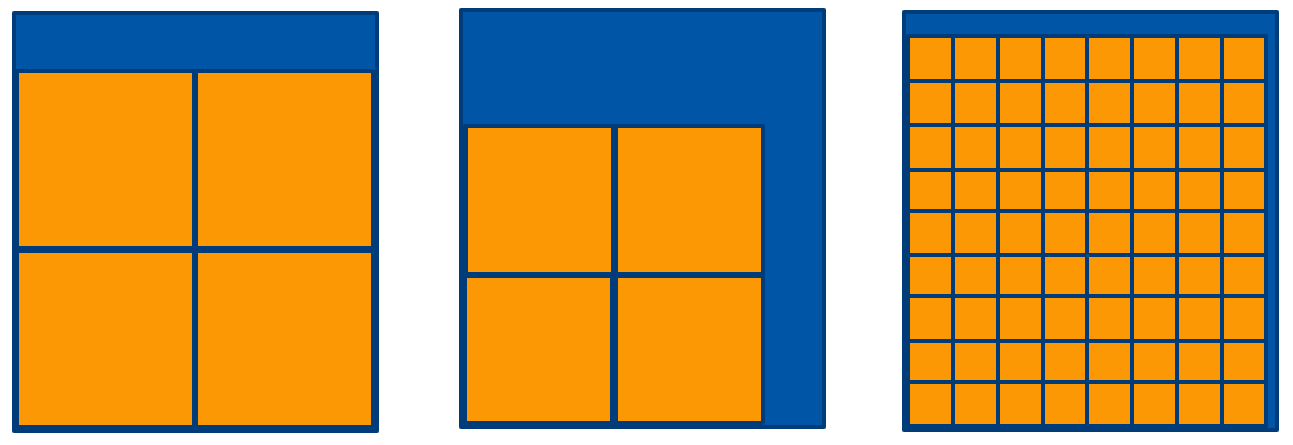}
                \caption{Reticle Utilization. In this example, square dies are placed in a reticle and the best utilization comes from the smallest dies.}
                \label{fig:reticle_utilization}
            \end{figure}

            Reticle utilization is the percentage of the reticle used by a chip. If the chip is much smaller than the reticle, many instances can fit in a single reticle. If the chip is large, it is more likely there will be a large waste of reticle area as shown in Figure \ref{fig:reticle_utilization}:\par

            \begin{equation}
                U_{reticle} = \frac{K_{reticle}A_{chip}}{A_{reticle}}
            \end{equation}

            \noindent where $K_{reticle}$ is the number of chips in a reticle.

            \begin{equation}
                K_{reticle} = \left\lfloor\frac{A_{reticle}}{A_{chip}}\right\rfloor
            \end{equation}

            Non-Recurring Engineering (NRE) costs are costs that occur once regardless of production volume and may be amortized over the production volume. We described this by splitting the NRE costs into design costs and mask costs.\par

            \begin{equation}
                \label{eq:nre_cost}
                C_{NRE}=\frac{C_{design}+C_{mask}}{N_{manufactured}}
            \end{equation}

            The NRE costs are amortized over the design volume $N_{manufactured}$. We split the design costs further into frontend and backend costs:\par

            \begin{equation}
                C_{NRE} = \frac{C_{BE} + C_{FE} + C_{mask}}{N_{manufactured}}
            \end{equation}

            \noindent where $C_{BE}$ and $C_{FE}$ are the backend and frontend design costs respectively.\par

            The mask costs are a sum of the mask costs for each layer and can be reduced by techniques such as multi-project wafers that share a reticle for multiple chip designs.\par

            \begin{equation}
                C_{mask}=P_{reticleshare}\sum_{i=1}^n C_{mask,i}
            \end{equation}

            $P_{reticleshare}$ is the percentage of the reticle that is taken by the design. In a standard design, this value will be 1, but it will can take a value smaller than 1 to model multi-project wafers and shared mask sets.\par

            The NRE design costs for both front end and back end vary by design type. Memories are very repetitive, so adding a memory to a design may only require running a generator script with the appropriate parameters. Designing logic will require much more thought from the designer including writing HDL and running place and route tools. Analog is the most intensive where typically both a logical design and a physical custom design are required. To account for this, the design is split according to the percentages of the core area that fall into each category.\par

            NRE frontend costs include the system definition and HDL development and testing. Backend includes physical design and implementation including cost of EDA tools for automated steps of the physical design:\par

            \begin{equation}
                C_{NRE,FE}=A_{core}(C_{L,FE}P_{L} + C_{M,FE}P_{M} + C_{A,FE}P_{A})
            \end{equation}

            \begin{equation}
                C_{NRE,BE}=A_{core}(C_{L,BE}P_{L} + C_{M,BE}P_{M} + C_{A,BE}P_{A})
            \end{equation}

            \noindent where $P_{L}$, $P_{M}$, and $P_{A}$ refer to the percentage of the design that falls into the logic, memory, or analog categories and each is multiplied by the appropriate cost for the category.\par
            
        \subsubsection{Die Yield}
\begin{figure}[t]
                \centering
                \includegraphics[width=0.5\columnwidth]{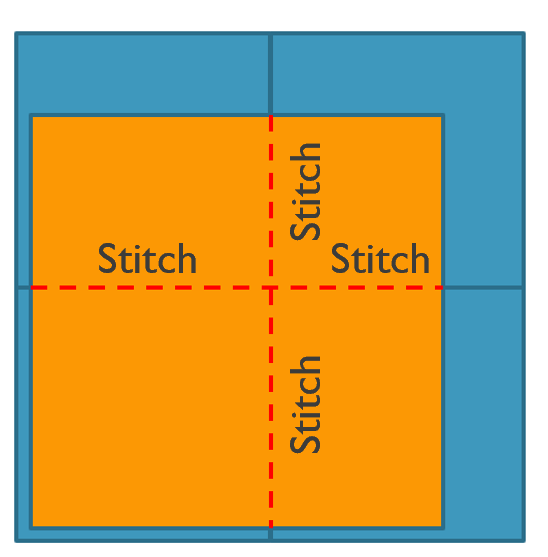}
                \caption{Reticle Stitching}
                \label{fig:reticle_stitching}
            \end{figure}

            Die yield is calculated based on the area of the chip and the negative binomial yield model based on the layer parameters. If there is more than one layer in the stackup, the yields of each layer are multiplied to get the total yield. If the chip is larger than a reticle, we also assume some stitching yield dependent on the number of times a chip crosses the reticle boundary shown in Figure \ref{fig:reticle_stitching}:\par

            \begin{equation}
                \label{eq:die_yield}
                Y_{die} = \prod_{i=1}^n Y_{layer,i}
            \end{equation}

            \begin{equation}
                \label{eq:layer_yield}
                Y_{layer} = Y_{stitching}\times Y_{defect}
            \end{equation}

            \begin{equation}
                \label{eq:defect_yield}
                Y_{defect} = (1+\frac{D_0A_C}{\alpha})^{-\alpha}
            \end{equation}

            \noindent where $D_0$ is the defect density and $A_C$ is the critical area determined as a product of the total area and the critical area ratio. The critical area ratio is the percentage of the area where defects will cause faults.\par

            \begin{equation}
                \label{eq:critical_area}
                A_C = (A_{core} + A_{IO})R_{critical}
            \end{equation}

            When a die is larger than a reticle, it is necessary to stitch multiple reticles together. To calculate the number of reticles required and the yield impact, we use the following equations.\par

            \begin{equation}
                \label{eq:stitching_yield}
                Y_{stitching} = Y_{PerStitch}^{K_{stitch}}
            \end{equation}

            \noindent where $K_{stitch}$ is the number of stitches.\par

            \begin{equation}
                N_{ret}=\left\lceil\frac{A}{A_{reticle}}\right\rceil
            \end{equation}

            We also need the number of stitches required. To do this, reticles are assumed to be placed in as close to a square configuration as possible then additional reticles are placed on the boundary.\par

            \begin{equation}
                N_{s,lsq}=\left\lfloor\sqrt{N_{ret}}\right\rfloor
            \end{equation}

            \noindent where $N_{s,lsq}$ is the number of reticles on a side of the largest square configuration of reticles lower than the number of reticles required. For dies that are only slightly smaller than a reticle, this will be one.\par

            \begin{equation}
                N_{lsq}=N_{s,lsq}^2
            \end{equation}

            \noindent where $N_{lsq}$ is the number of reticles in the square with $N_{s,lsq}$ reticles per side.\par

            \begin{equation}
                K_{stitch} = 2N_{s,lsq}(N_{s,lsq}-1)+2(N_{ret}-N_{lsq})-\left\lceil\frac{N_{ret}-N_{lsq}}{N_{s,lsq}}\right\rceil
            \end{equation}


            The final reported chip yield is based on the $Y_{self}$, which is the $Y_{die}$ for the die in question described above. This is multiplied by the assembly yield described in section \ref{sec:assembly_yield} and a quality yield term that depends on the expected percentage of stacked dies that are good.\par
            
            \begin{equation}
                Y_{chip}=Y_{self}Y_{assembly}Y_Q
            \end{equation}

            Quality yield is the yield determined by the probability that the chips in the assembly are good. In a scenario with perfect test, the quality yield would be 1, but in reality it will be smaller.\par

            \begin{equation}
                Y_Q=\prod_{i=1}^n Q_i
            \end{equation}

        \subsubsection{Assembly Cost}
            Assembly cost is primarily calculated as the cost of time spent on pick and place and bonding machines along with operator cost. Assembly machines can have multiple heads to simultaneously pick and place and bond dies, but there is variation in the parameters of the machines used. CATCH allows for setting both the size of the group of dies that can be placed simultaneously and the size of the group that can be bonded simultaneously. This allows for the user to set parameters to test methods ranging from individual pick and place followed by individual bonding as is frequently the case for thermal compression bonding \cite{sahoo2023} to simultaneous bonding using a method such as reflow. The cost of assembly is shown in equation \ref{eq:assembly_cost}:\par

            \begin{equation}
                \label{eq:assembly_cost}
                C_{assembly} = C_{PnP/s}T_{PnP} + C_{bond/s}T_{bond} + C_{mat}A
            \end{equation}

            \noindent where $C_{assembly}$ is the total cost of assembly, $C_{mat}$ is the cost of materials to bond per unit area, and $A$ is the area being bonded. The other parameters are cost and time parameters defined in the following equations.\par


            Pick and place time is computed simply as the number of dies to be bonded divided by the number of dies that can be placed at once as below.\par

            \begin{equation}
                T_{PnP} = \left\lceil \frac{N_{dies}}{G_{PnP}} \right\rceil t_{PnP}
            \end{equation}

            \noindent where $T_{PnP}$ is the pick and place time, $t_{PnP}$ is the time for a single pick and place step, $N_{dies}$ is the number of dies to be bonded, and $G_{PnP}$ is the size of the group that can be placed simultaneously.\par

            Bonding time is computed simply as the number of dies to be bonded divided by the number of dies that can be bonded at once as below.\par

            \begin{equation}
                T_{bond} = \left\lceil \frac{N_{dies}}{G_{bond}} \right\rceil t_{bond}
            \end{equation}

            \noindent where $T_{bond}$ is the total bonding time, $t_{bond}$ is the time for a single bonding step, and $G_{bond}$ is the size of the group that can be bonded simultaneously.\par

            The cost of assembly is computed by multiplying the time required by the cost per second.\par

            \begin{equation}
                C_{assembly} = T_{bond}*K_{bond} + T_{PnP}*K_{PnP}
            \end{equation}

            \noindent where $K_{bond}$ and $K_{PnP}$ are the cost of time spent in bonding and pick and place respectively.\par






        
        \subsubsection{Assembly Yield} \label{sec:assembly_yield}
            Assembly yield is the yield of stacked top \textit{Chips} on the bottom \textit{Chip}. This is dependent on many factors. For simplicity, we break it down into three main components. The first, bonding yield, is the probability that an individual bond succeeds. This scales according to the number of pins bonded. The second component of yield is the die alignment. This scales with the number of dies. The third component is included for hybrid bonding. In hybrid bonding, there is a need to have a very clean bonding surface for the initial dielectric bond. Even a very small particle can leave a large region where the bond fails due to the rigidity of silicon. This can interfere with the connections formed during annealing \cite{nagano2022}. To model this factor, we added a term for the probability that there are no particles on the bonding surface given a defect density and bonding area. This term is a reduced form of the negative binomial yield model with a clustering factor of one.\par

            \begin{equation}
                Y_{assembly} = Y_{bond}^{N_{pin}} \times Y_{alignment}^{N_{dies}} \times \frac{1}{1 + D_{H}A_{bond}}
            \end{equation}

            \noindent where $Y_{assembly}$ is the assembly yield, $Y_{bond}$ is the yield per pin, $N_{pin}$ is the number of pins to bond, $Y_{alignment}$ is the alignment yield for a single die, and $N_{dies}$ is the number of dies. $D_{H}$ is the defect density per unit area on the dielectric used for hybrid bonding and $A_{bond}$ is the area of the hybrid bond. The defect density for the dielectric bond can be set to zero if hybrid bonding is not used.\par

    
            
        


    \subsection{Test}

        Splitting a monolithic design into multiple chiplets has mixed impacts on the testing process. Smaller chips provide the opportunity for more fine-grained tests. This can improve the quality of the final product \cite{Singh1994}. Although the chips will be better tested, testing time goes up since chips will be tested individually before being assembled on the interposer and likely again after assembly to ensure assembly did not introduce any errors. To capture this, we include a test cost model in CATCH.\par

        We assume a scan-chain style test. The scan chain length corresponds to the number of cycles of the test clock required to load a test pattern. The testing cost is based on the time required for test and the cost of time spent on the testing machine. As calculating the number of test patterns required to achieve a certain fault coverage is design specific, we take the required number of patterns and the scan chain length as inputs along with the test clock period and the cost of time spent on the testing machine.\par

        \begin{equation}
            C_{test} = K_{machine}N_{patterns}L_{sc}T_{test}
        \end{equation}

        \noindent where $K_{machine}$ is the cost per second of operating the test machine, $N_{patterns}$ is the number of test patterns required to achieve the defined fault coverage, $L_{sc}$ is the scan-chain length (corresponding to the number of clock cycles per test), and $T_{test}$ is the period of the test clock in seconds.\par

        \begin{equation}
            Y_{test} = 1 - K_{coverage}(1 - Y_{true})
        \end{equation}

        \noindent where $K_{coverage}$ is the fault coverage and $Y_{true}$ is the yield calculated as described in previous sections.\par
        
        Since testing may be performed at different steps of assembly in a chiplet-based system, testing parameters are defined separately for individual dies and for the stacked assembly at each stage. This provides flexibility to evaluate more thorough (and expensive) tests at different steps of the process.\par

    \subsection{Yield/Quality/Cost Structure}

\begin{figure*}[t]
            \centering
            \includegraphics[width=1.8\columnwidth]{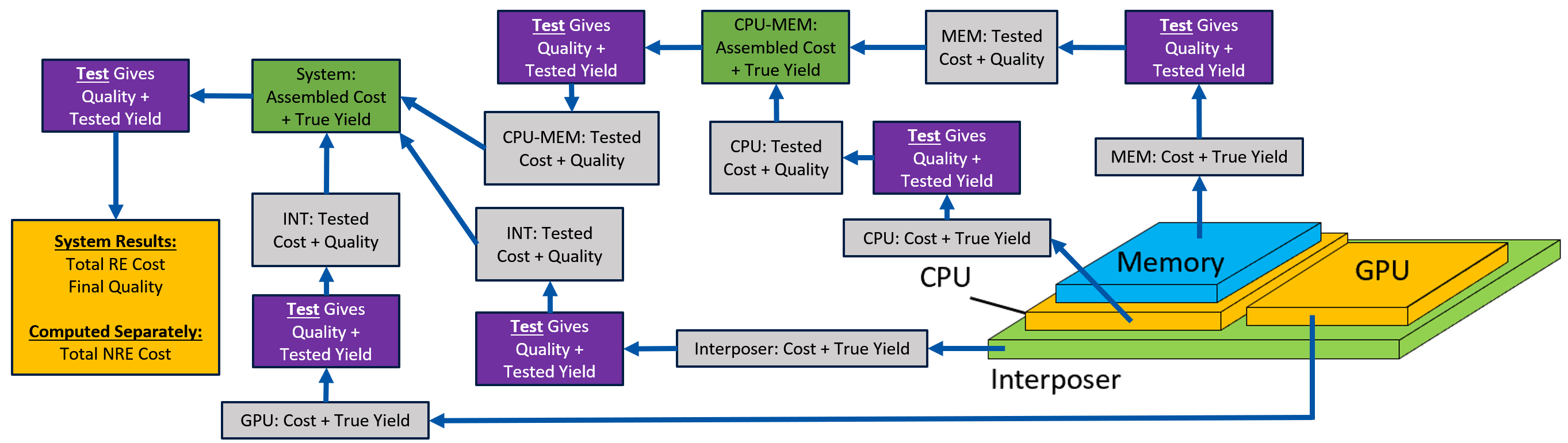}
            \caption{Yield and Cost Propagation}
            \label{fig:yield_cost_propagation}
        \end{figure*}

        To handle the cost and yield of the system and the components recursively, we determine two values for each component \textit{Chip}: quality and tested cost. Quality is the percentage of post-test chips that are expected to be good. Tested cost is the cost accounting for the yield of the system.\par

        \begin{equation}
            C_{tested} = \frac{C_{untested}}{Y_{test}}
        \end{equation}

        \noindent where $C_{untested}$ is the cost of the \textit{Chip} itself along with the tested costs of any component upper \textit{Chips} and cost of assembly. The $Y_{test}$ is the percentage of the assembled \textit{Chip} plus upper \textit{Chips} that will pass test. The cost and yield propagation is shown in Figure \ref{fig:yield_cost_propagation}. Note that the NRE costs are computed in parallel and added to the final RE costs since NRE is independent from the yield.\par

\section{Tool Interface/Overview}

    The tool is designed to be user friendly and fully parameterized. This is accomplished by using a multi-file structure for parameter and system definitions. The input files follow an XML format.\par

    \subsection{File Structure}

        \subsubsection{System Description}

            The system description consists of two files: a \textit{Chip} definition file and a \textit{Chip}-level netlist. These files change with different systems.\par
            
            The \textit{Chip} definition file contains the parameters that are specific to each \textit{Chip} in the stack along with the names of processes defined in the library file for testing, assembly, layers, and wafer process. If one \textit{Chip} is stacked on another \textit{Chip}, the upper \textit{Chip} definition is nested in the lower \textit{Chip} definition.\par

            The netlist simply contains a reference to the IO type and the names of the \textit{Chip} objects which are connected by that IO. To define external connections, simply set one of the ends of the connection to a chip that does not exist in the \textit{Chip} definition file. This can be a dummy name like ``external'' or the name of a block which you are not including in the analysis. In this case, only one of the ends of the IO will count toward the area of the system.\par

        \subsubsection{Library Files}

            The process names referenced in the \textit{Chip} definition file are contained in a set of library files. These are expected to be reused across studies.\par

            The IO definition file contains a list of defined IO types which may be selected from in the netlist file. Each IO has the option to define a different area for the \textit{from} and \textit{to} cells. In the case the IO is bidirectional, these should be the same. Each IO also has a maximum bandwidth which it can support, a number of wires/pads required for the protocol, a maximum reach, and an energy per bit.\par

            The assembly process definition file contains information on the cost of time spent on the bonding machine along with the time required for a single step of pick and place and for a single step of bonding. Since multiple dies may be placed at a time or multiple dies may be bonded simultaneously depending on the machine, there is a bonding group parameter as well. In the case of wafer to wafer bonding, bonding group and pick and place group should be set large enough to fit the entire wafer of dies. The assembly process definition also includes the die separation, edge exclusion for stacking one die on another, maximum current density of the pads, and the yield parameters for estimating the yield of the assembly step.\par

            The test process definition includes time spent per test cycle, cost of time spent on the test machine, pattern count, scan chain length, and defect coverage. This is included for both self-test and assembly-test, so different amounts of test may be allocated to different steps of the assembly process.\par

            Layers are defined based on a cost per $mm^2$. This should be the cost assuming the wafer is fully utilized with no waste. The layer definition also contains yield parameters for the negative binomial yield model, percentage of layer cost that is due to time in lithography, mask cost, and stitching yield.\par

            Defect densities are derived from the numbers in \cite{Shilov2024}. Since we do not have comprehensive numbers for defect densities for each technology node, we assume older technologies have equal or better defect densities than newer. We also use the number for defect density at manufacturing unless otherwise stated.\par



\section{Case Studies} 

    \subsection{Test Case}

\begin{figure}
            \centering
            \includegraphics[width=0.9\columnwidth]{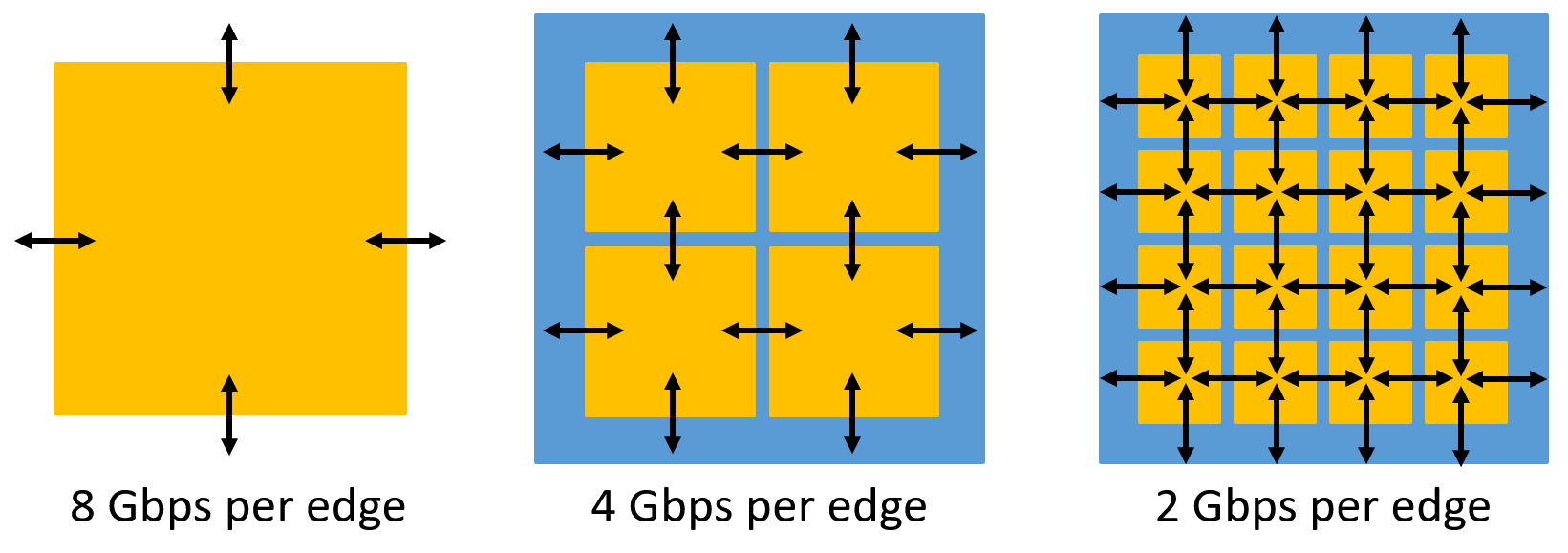}
            \caption{Homogeneous Graph Processor. The graph processor is shown for the monolithic case and for 4 or 16 chiplets. 8 Gbps per edge in the monolithic case is an example value. Actual number is set for each case study. The marked edge bandwidths ensure that the external bandwidth is the same in each case.}
            \label{fig:homogeneous_graph_processor}
        \end{figure}

        In the following case studies, we used a test case based on a generic graph processor. Our generic graph processor is a homogeneous chiplet design. This is the test case used in \cite{Graening2023}. For this test case, we assume we have an 800 $mm^2$ chip. Since this is smaller than a standard reticle (for a 33$\times$26 mm reticle, area is 858 $mm^2$) it can be manufactured as a monolithic die if yield is reasonable. We consider splitting this theoretical graph processor chip into different numbers of chiplets. Since this is a homogeneous case, we consider that the edge bandwidth is constant as shown in Figure \ref{fig:homogeneous_graph_processor}. As CATCH does not require circuit specifics, we assume that the number of cores in the graph processor divides evenly by the number of chiplets we use to implement the design and that each chiplet is identical.\par

    \subsection{Default Assumptions}

        The default assumptions regarding values used for the case studies are shown in Table \ref{tab:assumptions}. The defect density is a good defect density for a mature process \cite{Shilov2024}. The critical area ratios chosen were used to provide some yield scaling between the technology nodes as similar sized defects will be less likely to cause critical faults in less advanced nodes. Silicon cost per mm$^2$ is from the 300 mm wafer costs in \cite{Shilov2020}, but note that the actual cost of dies will vary based on wafer utilization and yield.\par
        
        \begin{table}[t]
            \centering
            \caption{Table of Assumptions}
            \begin{tabular}{|l|r|r|r|r|r|r|}
                \toprule
                      & \multicolumn{1}{l|}{3nm} & \multicolumn{1}{l|}{5nm} & \multicolumn{1}{l|}{7nm} & \multicolumn{1}{l|}{10nm} & \multicolumn{1}{l|}{12nm} & \multicolumn{1}{l|}{40nm} \\
                \midrule
                Defect Den. (/cm$^2$) & 0.5   & 0.5   & 0.5   & 0.5   & 0.5   & 0.5 \\
                \midrule
                Critical Area Frac. & 0.7   & 0.67  & 0.64  & 0.62  & 0.6   & 0.5 \\
                \midrule
                Cost per mm$^2$ & 0.29  & 0.25  & 0.13  & 0.085 & 0.056 & 0.034 \\
                \bottomrule
            \end{tabular}%
            \label{tab:assumptions}%
        \end{table}%

    \subsection{Cost Optimal Chiplet Size}

        The cost-optimal size for chiplets in a chiplet system depends on many factors. In this case study, we look at the homogeneous graph processor test case and sweep the number of chiplets from 4 to 64.

        \subsubsection{Technology Node}
        
\begin{figure}[t]
                \centering
                \includegraphics[width=0.9\columnwidth]{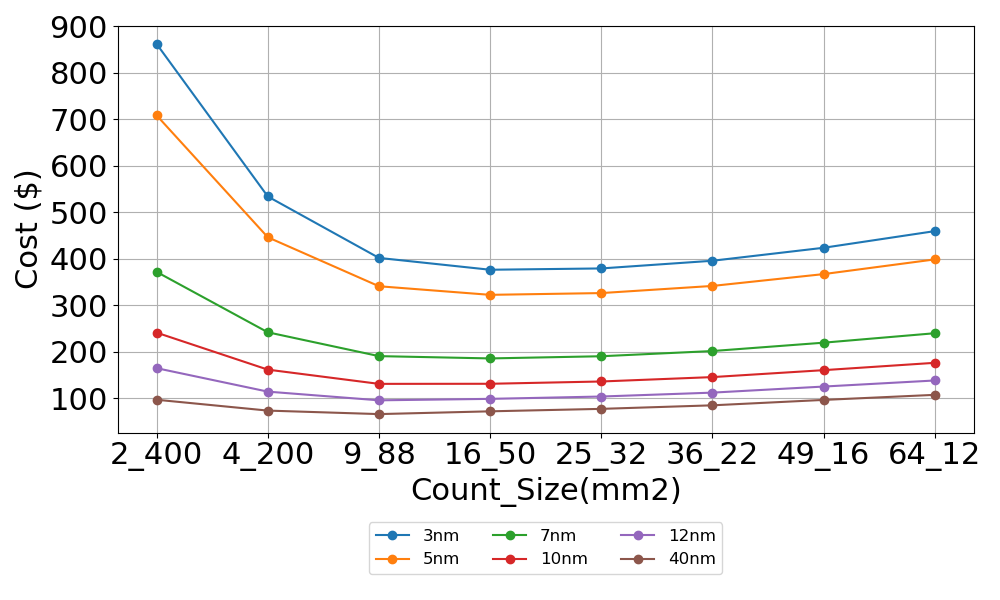}
                \caption{Optimal Chiplet Size Across Technologies. The best shown chiplet size configuration for 3nm is 9, 88 $mm^2$ chiplets and the best case for 40nm is 4, 200 $mm^2$ chiplets.}
                \label{fig:size_sweep}
            \end{figure}

            The cost-optimal chiplet size varies depending on the technology node. The results of running a sweep in multiple technology nodes are shown in Figure \ref{fig:size_sweep}.\par

        \subsubsection{Process Node Maturity}

\begin{figure}[t]
                \centering
                \includegraphics[width=0.9\columnwidth]{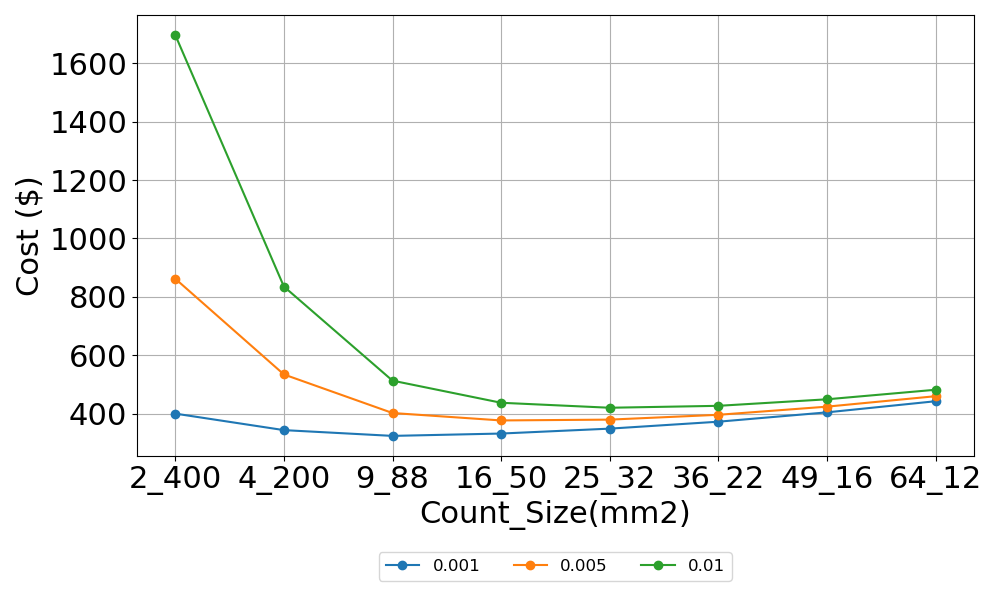}
                \caption{Optimal Chiplet Size Across Defect Densities. The best shown chiplet size configuration is different for each defect density.}
                \label{fig:defect_density}
            \end{figure}

            The defect density of immature process nodes will be higher than it will be for mature processes. Since this significantly impacts yield, it will also have an impact on the optimal chiplet size. In Figure \ref{fig:defect_density}, three different defect densities are shown for the 3nm node. Defect density of 0.005 per mm$^2$ corresponds to the defect density used in Figure \ref{fig:size_sweep} and the other studies except where noted.\par

        \subsubsection{Assembly Type}

\begin{figure}[t]
                \centering
                \includegraphics[width=0.9\columnwidth]{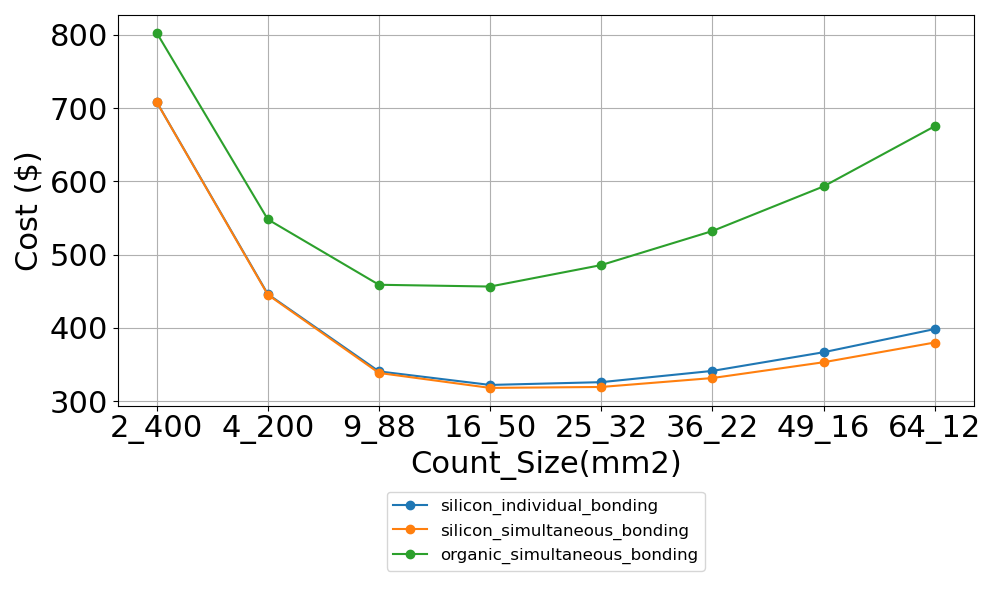}
                \caption{Assembly Process Sweep. X-axis shows the number of chiplets and the size of each chiplet used for the study.}
                \label{fig:assembly_sweep}
            \end{figure}

            Assembly cost depends largely on time spent on the assembly machine. This can be impacted by the assembly strategies regarding placement and bonding. Typical PCB assembly involves placing each component individually, then using solder reflow as a bonding step. More advanced bonding techniques may have a process where dies are bonded individually in addition to being placed individually. Different assembly strategies will begin to have a substantial impact on cost for larger numbers of chiplets. In Figure \ref{fig:assembly_sweep}, we compare the scaling for a silicon substrate using individual bonding with a silicon substrate using simultaneous bonding. The IO type is UCIe standard for the organic case and UCIe advanced for the silicon cases. In the results shown here, organic is more expensive than silicon mostly because the UCIe standard IO cell is larger and supports a lower bandwidth than the UCIe advance IO cell mostly due to bump pitch. At the end with the larger number of chiplets, you also see some impact due to the cost of assembly.\par

        \subsubsection{Non-Recurring Engineering Cost}
        
\begin{figure}[t]
                \centering
                \includegraphics[width=0.9\columnwidth]{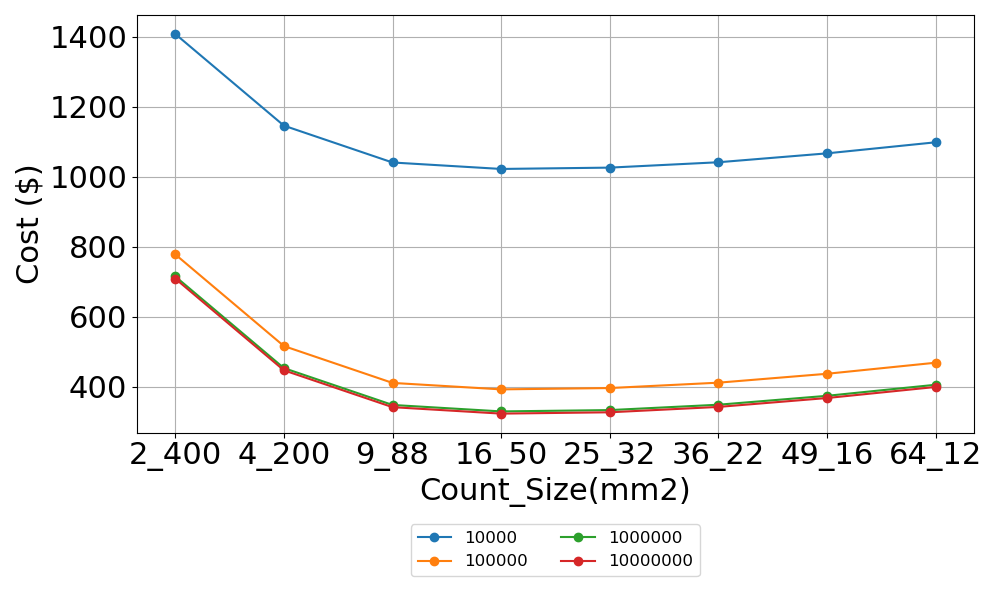}
                \caption{NRE Sweep. Design cost was included and scaled based on the estimated amount of duplication in each die.}
                \label{fig:nre_sweep_design}
            \end{figure}

            Non-Recurring Engineering (NRE) costs include costs such as masks and tooling that occur regardless of manufacturing volume. In most of the case studies, we have assumed a high manufacturing volume so these costs are small. For low volume manufacturing with non-homogeneous chiplet designs, NRE costs are a significant factor making smaller chiplets less cost-effective (due to a greater number of masks) unless chiplets can be designed to be reusable across a wide range of designs. In Figure \ref{fig:nre_sweep_design}, we show how the cost decreases as manufacturing quantity increases for our homogeneous test case. The shape of the chiplet configuration does not change between quantities since the quantity scales with the number of dies on the system as all dies are the same in this test case. As an approximation, since this system is approximately reticle sized, regardless of the die size chosen here, all dies for one assembly approximately fit in a single reticle. Additionally, since this is a homogeneous graph processor case, we assume the same design costs go into the system regardless of number of chiplets as would be the case if a single block was designed and copied across the rest of the system.\par


    \subsection{Test Cost}

\begin{figure}
            \centering
            \includegraphics[width=0.9\columnwidth]{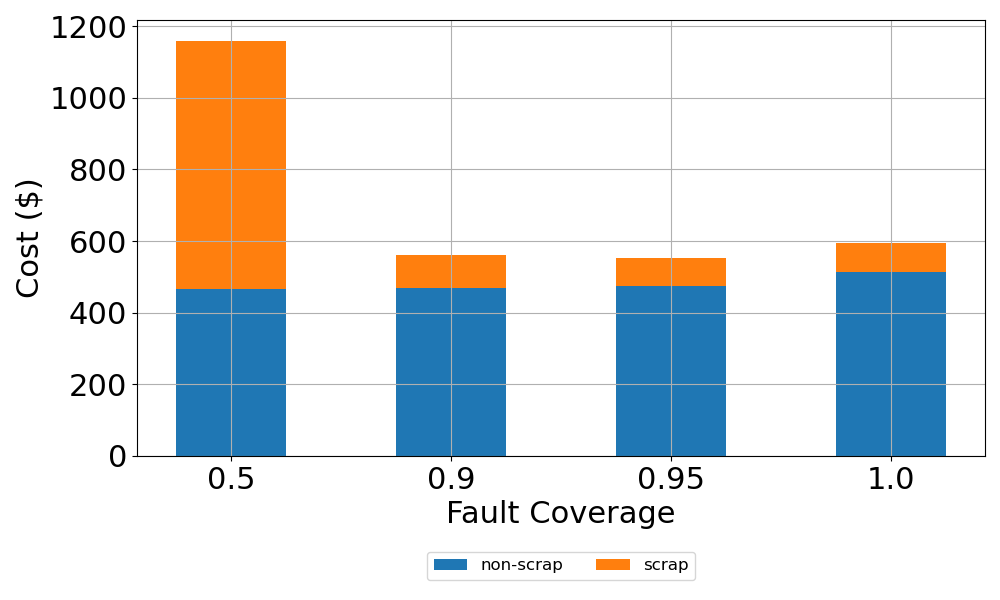}
            \caption{Fault Coverage Sweep for Mature Process Node.}
            \label{fig:test_sweep}
        \end{figure}

\begin{figure}
            \centering
            \includegraphics[width=0.9\columnwidth]{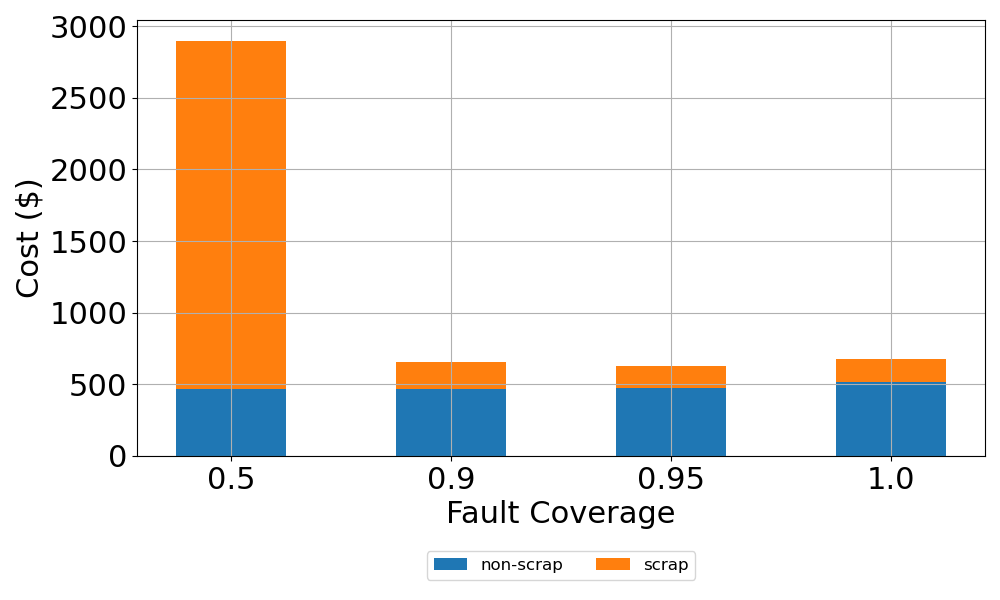}
            \caption{Fault Coverage Sweep for Immature Process Node. Due to the increase in test cost for the higher fault coverage, the best cost in this case is the 95\% fault coverage. 50\% fault coverage is the worst due to extremely high scrap cost.}
            \label{fig:test_sweep_immature}
        \end{figure}

        Fault coverage at intermediate testing steps impacts the stack yield and scrap cost associated with bad dies. If the test coverage is too low, good dies are wasted when assemblies including good dies are thrown away because a bad die was not detected during test. The fault coverage does come at the expense of increased test costs. This is shown in Figure \ref{fig:test_sweep} and \ref{fig:test_sweep_immature} for the graph processor test case. We used the graph processor configuration with 16, 50 mm$^2$ chiplets for these results and used the 3nm node with both the elevated defect density of 0.01 per mm$^2$ and the good defect density for mature processes of 0.005 per mm$^2$. Due to the number of chiplets, poor testing at the die level results in many of the assembled systems failing, giving high scrap cost. Dies are tested individually and after each assembly step. For the 2.5D case, this means the dies are tested before assembly and once all dies are on the interposer.\par
        
        
        
        


\begin{figure}
            \centering
            \includegraphics[width=0.9\columnwidth]{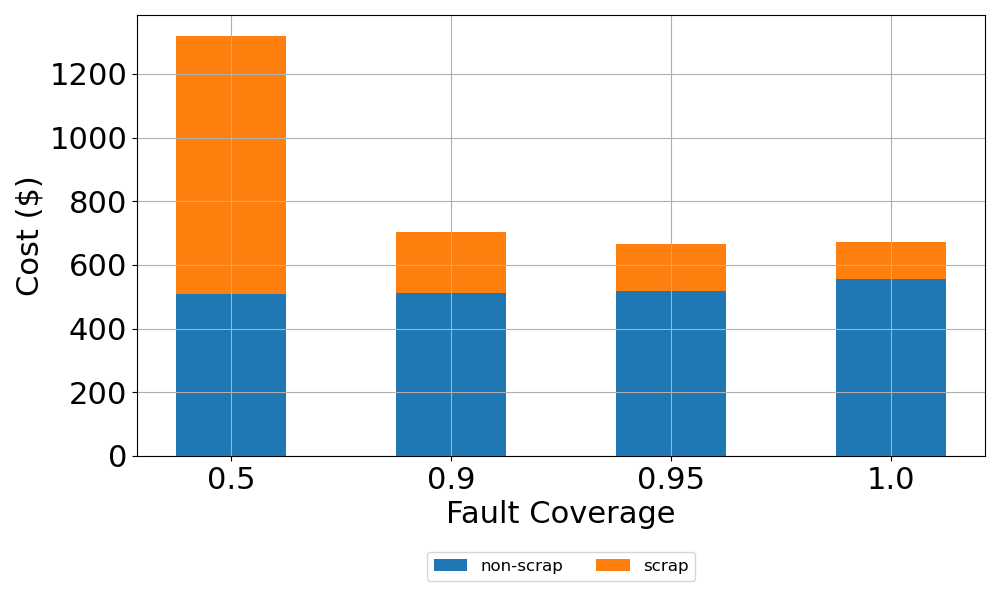}
            \caption{Fault Coverage Sweep for 3D Case in Mature Process Node.}
            \label{fig:test_sweep_3d}
        \end{figure}

\begin{figure}
            \centering
            \includegraphics[width=0.9\columnwidth]{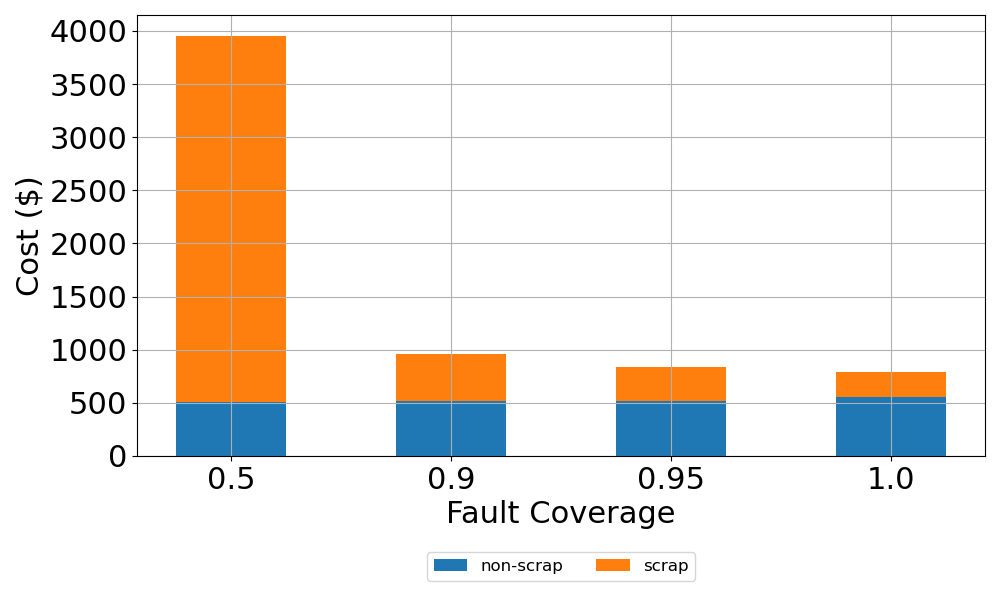}
            \caption{Fault Coverage Sweep for 3D Case in Immature Process Node.}
            \label{fig:test_sweep_3d_immature}
        \end{figure}

        The results change if we instead use a 3D stacked system instead of a 2.5D integrated system. Where the first case included 16 dies placed beside each other on an interposer, in this case, we stacked the 16 dies one on top of the other. We assume die-to-wafer stacking so that individual dies are only stacked if the rest of the stack has tested as functional. These results are shown in Figure \ref{fig:test_sweep_3d} and \ref{fig:test_sweep_3d_immature}. In this case, the increased testing gives better yield and lower scrap cost than the 2.5D case. \par

    \subsection{IO Reach}

\begin{figure}[t]
            \centering
            \includegraphics[width=0.9\columnwidth]{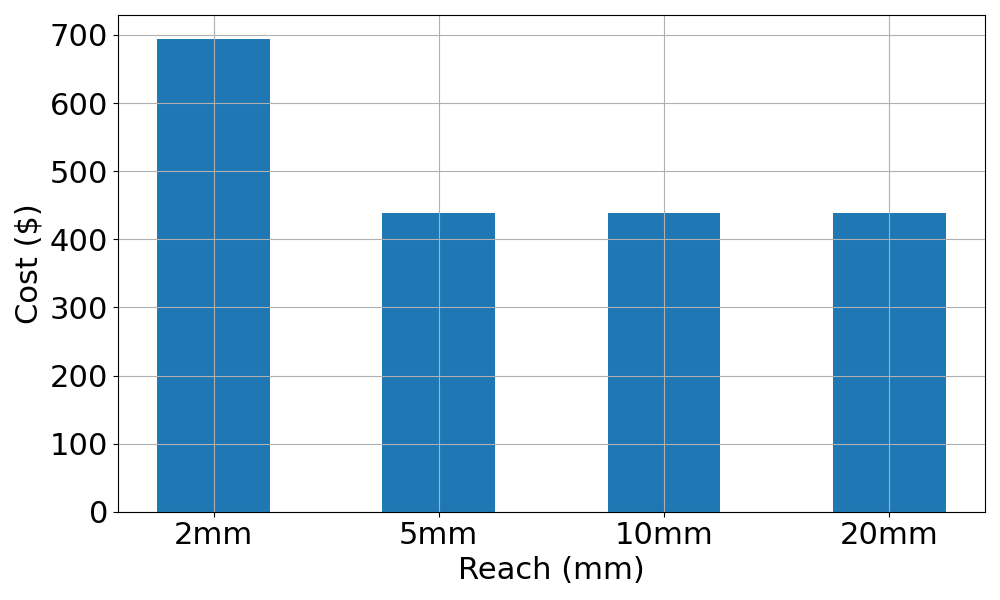}
            \caption{Reach Sweep.}
            \label{fig:reach_sweep}
        \end{figure}

        IO reach limits the region on the die where an IO cell can be placed. In an IO constrained example, this can require increasing the die size to increase the perimeter where the IO cell can be placed. To demonstrate this, Figure \ref{fig:reach_sweep} shows how cost changes with respect to the reach of the IO cell. For this study, we used a parallel IO type and scaled the reach instead of using UCIe as was used for the other studies. Note that the 2mm reach allows mostly edge placement of IO cells, where the longer reaches allow for placement of IO cells further from the edge of the die. We see a penalty in the die area when the reach is too small since the die size is increased to provide more perimeter for edge placements of IO cells, but when the reach is large enough, we no longer see a benefit to scaling the reach.\par

    \subsection{Interposer Study}

\begin{figure}[t]
            \centering
            \includegraphics[width=0.9\columnwidth]{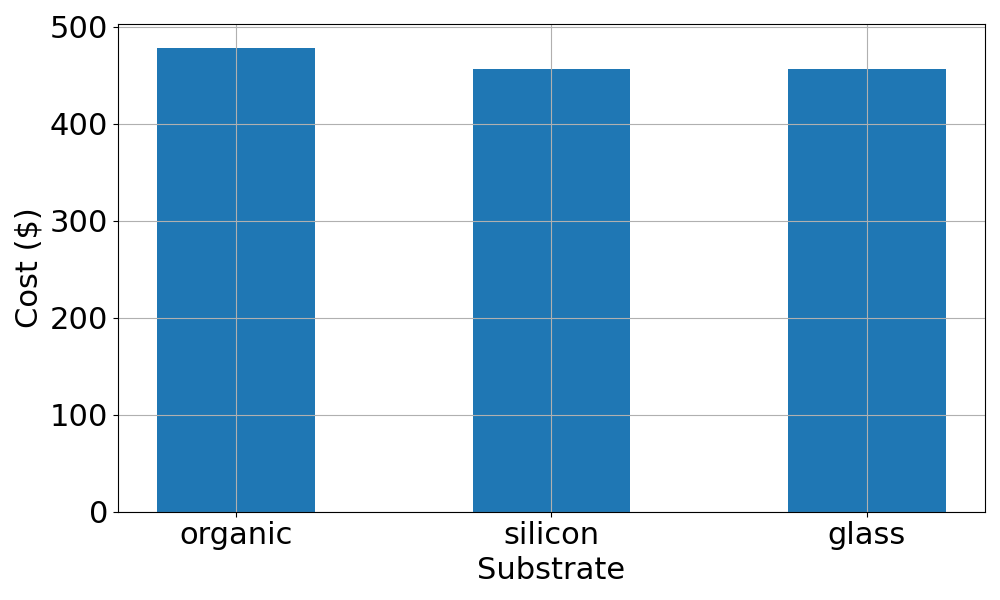}
            \caption{Interposer Material Comparison.}
            \label{fig:substrate_sweep}
        \end{figure}

        Chiplet systems using 2.5D stacking require some kind of substrate or interposer between dies and the PCB. Different substrate materials impact the achievable bonding pitches, cost, and yield. Silicon substrates have the advantage of sharing the coefficient of thermal expansion (CTE) of the dies which allows use of extremely small bonding pitch. This comes at the disadvantage of being a more expensive option than using an organic substrate. An additional consideration is that organic substrates tend to have lower loss than silicon. Glass interposers try to take some of the best of both worlds with smaller pitch than organic substrates and better signal integrity than silicon. Organic, silicon, and glass substrates are compared on the homogeneous test case for a 16-chiplet design in Figure \ref{fig:substrate_sweep}. The IO type is UCIe standard for the organic case and UCIe advanced for the silicon and glass cases.\par

        Another integration option is to use an embedded bridge die that allows high density interconnections through the silicon, but allows the body of the substrate to be a material other than silicon to reduce cost and improve yield. To support this option in the cost model, we allow setting a die as ``buried'' which will interpret it as not taking up area on the surface of the interposer as it will be place under the dies that are bonded on top and level with the interposer surface.\par

    \subsection{Non-Recurring Engineering Cost}
        
        One useful applications for chiplets is the situation where you want to update one die in an existing system. This can allow reusing most of the dies, but adding one a new die for additional functionality or a different application. The primary new cost in this situation is the increased NRE cost of designing and manufacturing a new die.\par
        
\begin{figure}[t]
            \centering
            \includegraphics[width=0.9\columnwidth]{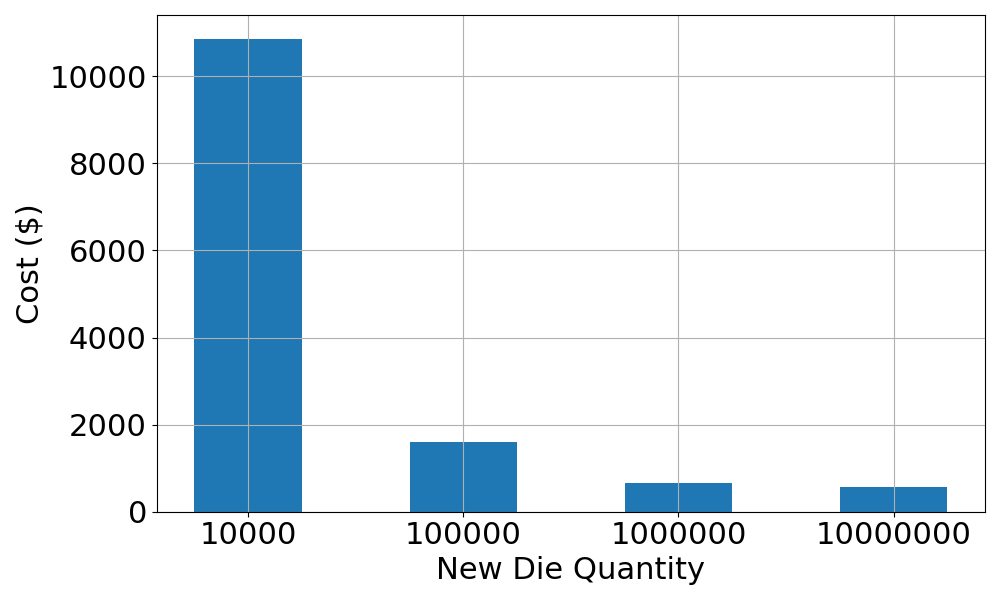}
            \caption{NRE Impact with One New Die. The cost of the existing high-volume dies is almost negligible for very low volumes.}
            \label{fig:nre_sweep_1_die}
        \end{figure}

        To see the impact of NRE cost on a system like this where one die is new and low volume while all the others are high volume either due to reuse across a large number of designs or because they were previously designed for a different system, we looked at a 4-chiplet system using the homogeneous test case, but varying the quantities. In this case 3 of the 4 dies were set to a high enough quantity to make NRE negligible, then the quantity of the other die was varied. This is shown in Figure \ref{fig:nre_sweep_1_die}. Here despite the new die only making up 25\% of the silicon area, NRE cost dominates.\par

\section{Conclusion} \label{sec:conclusion}

    In this work, we have shown that CATCH is a valuable tool to analyze chiplet systems at an early design stage using high level design descriptions. Our tool includes detailed modeling for cost and yield for silicon, assembly, and test. Additionally, we include netlist support to show the impact of IO cells and inter-chiplet bandwidths. The tool allows exploring a wide variety of options to make informed decisions about the physical configuration of a chiplet system during design and it helps identify important areas of high cost where improvement will have the largest impact in system cost. Since this is based on a flexible design description format it is also well-suited to integrations with other tools for co-optimizations in future work.\par






%


\bibliographystyle{IEEEtran}
\bibliography{references, bstcontrol, IEEEabrv}

\end{document}